# Atmospheric conditions and their effect on ball-milled magnesium diboride

B J Senkowicz[1,3], R Pérez Moyet[2], R J Mungall[1], J Hedstrom[1], O N C Uwakweh[2], E E Hellstrom[1], and D C Larbalestier[1,3]

[1]Applied Superconductivity Center and Department of Materials Science and Engineering, University of Wisconsin – Madison, Madison, WI 53706

[2]University of Puerto Rico Mayaguez

[3]Now at Florida State University and the National High Magnetic Field Laboratory – Tallahassee, Florida

E-mail: bjsenkowicz@wisc.edu

**Abstract.** Magnesium diboride bulk pellets were fabricated from pre-reacted $MgB_2$ powder ball milled with different amounts of exposure to air. Evidence of increased electron scattering including increased resistivity, depressed $T_c$, and enhanced $H_{c2}$ of the milled and heat treated samples were observed as a result of increased contact with air. These and other data were consistent with alloying with carbon as a result of exposure to air. A less clear trend of decreased connectivity associated with air exposure was also observed. In making the case that exposure to air should be considered a doping process, these results may explain the wide variability of "undoped" $MgB_2$ properties extant in the literature.

**1. Introduction**

$MgB_2$ technology has advanced rapidly since its discovery as a superconductor in 2001.[1-4] Long lengths of wire and tape[5-7] have been fabricated by both the in-situ and ex-situ methods, and several >1T coils have been wound.[8-13] Local $J_c$(8 T, 4.2 K) values in bulk samples are beginning to approach those of NbTi wires.[14-18] It remains unclear whether the in-situ or ex-situ wire is preferable, given that they have different advantages.

Despite rapid advances, $MgB_2$ exhibits several experimental peculiarities. The most notable of these may be the broad range of intragranular properties including $T_c$, normal state resistivity ($\rho$), and $H_{c2}(T)$ found for nominally pure material synthesized by different methods.[19-23] The most spectacular variation has been in measured $H_{c2}$ values. The undoped samples of Bud'ko et al.[21] had $H_{c2}(0\ K) \sim 16$ T, in contrast to the $H_{c2}(0\ K)$ value of 20.5 T reported for an undoped sample by Bhatia et al.[22], and $H_{c2}(10K)$ ranging from 17 T to 24 T quoted for undoped sample by Matsumoto et al.[28]. In the rush to understand carbon and SiC doping, not enough attention has been devoted to understanding the wide variation in intragrain properties between nominally undoped samples. Some seemingly uncontrolled mechanism has a substantial effect on these properties. In some samples, particularly those heat treated at low temperatures, a high concentration of microstructural defects may increase electron scattering. A more widely applicable explanation for the dependence of properties on processing conditions is that most $MgB_2$ is contaminated (in some cases quite heavily) due to reaction of the precursor powders with air.[24] This experiment examines the effect of atmospheric contact before and after ball milling commercially available (Alfa-Aesar) pre-reacted $MgB_2$ precursor powder by examining the properties of sintered bulk samples subsequently made from the milled powder.

## 2. Experimental Procedures

All four samples in this set were made from Alfa-Aesar pre-reacted $MgB_2$. We milled for 10 hours in a Fritsch Pulverisette 4 ball mill both to promote compositional homogeneity and in order to facilitate interaction with gas in the milling jar. For all samples, the WC milling jar was housed inside a specially made brass can with a compressed O-ring seal and a port for an optional valve used only for sample D. Milling was carried out with hardened steel balls. Although there is a possibility of contamination by the milling apparatus, the jar and balls did not show any visible wear, and there was no evidence for iron or WC contamination based on in x-ray diffraction patterns. Moreover the milling procedure was identical for each sample.

The four samples were handled and milled under slightly different atmospheric conditions. For sample A, the milling jar was loaded and sealed into the brass can in a glove box filled with dry nitrogen and equipped with active oxygen and water scavenging. After milling, the jar was opened and the powder was handled in the same glove box. For sample B, a similar milling jar was loaded and sealed in the brass can in a nitrogen filled glove bag filled with flowing $N_2$ gas from which residual oxygen and water were not removed. Unsealing and post-milling handling were done in the glove box, as for A. For samples C and D, the milling jars were loaded and sealed in the brass can in the typically humid atmosphere of Mayaguez, Puerto Rico and, after milling, were unsealed and handled in the dry winter air in Madison, Wisconsin. The brass can was hermetically sealed for sample C, whereas for sample D the brass can was fitted with a one-way check valve that could open to vent excess pressure from the milling can. The handling atmospheres and property summary are listed in Table I.

The milled powders were pressed into pellets and sealed in evacuated stainless steel tubes, then hot isostatic pressed at $1000^{o}C$ and ~200 MPa for 5 hours. The pellets were sufficiently dense after sintering that they had hardness > 10 GPa and needed to be sectioned with a diamond saw before measurement.

3. Results

X-ray diffraction patterns are shown in Fig. 1 using a linear intensity scale. All four samples had similar XRD patterns, and all contained small amounts of MgO and $MgB_4$. The peak at ~$63^{o}$ is a composite of an $MgB_2$ peak at $63.27^{o}$ and an MgO peak at $62.22^{o}$. For samples A, B, C, and D the peak positions are $62.85^{o}$, $62.8^{o}$, $62.2^{o}$, and $62.3^{o}$ respectively. The shift to lower $2\theta$ for samples C and D indicated that those samples milled in air contained more MgO than samples A and B, which were milled in an inert atmosphere.

The superconducting moment was measured in a SQUID magnetometer as a function of increasing T after zero-field-cooling to 5 K. Figure 2 shows the magnetic moment **m** as a function of temperature between 36 K and 39 K, normalized to the moment at 5 K. Sample A had the highest $T_c$ and the sharpest transition. Sample B had an onset $T_c$ nearly as high as sample A, but its transition was not as sharp. Sample C had a sharp transition, beginning ~0.1 K below that of sample A. Sample D also had a fairly sharp transition, beginning ~0.5 K below that of sample A. Thus there was a small monotonic depression of Tc with increasing exposure to air.

Using a Quantum Design Physical Property Measurement System (PPMS) we measured the normal state resistivity as a function of temperature and field in the normal and in the superconducting state.

Figure 3 shows the dependence of the upper critical field $H_{c2}$ on temperature, defined using 90% of the normal state resistivity as our criterion for $H_{c2}$. The trend in $H_{c2}$ was similar to that seen in $T_c$ but of opposite sign – as atmospheric exposure increased, $H_{c2}$ increased. The $H_{c2}$ values at 24K increased from 7.0 T for sample A to 7.9 T for sample D. Our data fall between $H_{c2}$ values from the literature taken on pure and 3.8% carbon-doped filaments.[30]

The temperature dependence of $\rho$ is shown in Fig. 4. We observed an increase in normal state resistivity at all temperatures, as well as an increase in $\Delta\rho = \rho(300\ K) - \rho(40\ K)$ and a decrease in RRR from samples A through C corresponding to the $MgB_2$ having had increasing exposure to air. Sample D falls between samples B and C in measured $\rho$ and in $\Delta\rho$, but has the lowest RRR of the set.

Fig. 5 shows $J_c(H)$ at 4.2 K for each sample, as calculated from M-H hysteresis loops using the equation $J_c(H,T) = 0.5*\Delta M*12b/(3bd-d^2)$, where b and d are the width and thickness of the rectangular section bar and $\Delta M$ is the difference in volumetric magnetization between the arms of the MH loop.[32] Low field $J_c$ was very similar for all samples, but high field $J_c$ increased progressively from A to D, and $H^*(J_c = 100A/cm^2)$ ranged from less than 8 T for sample A to more than 9 T for sample D.

## 4. Discussion

The Alfa Aesar $MgB_2$ used for this experiment is one of the most widely used starting materials for experiments on pre-reacted $MgB_2$ yet our post-processing resulted in clear property change trends in every quantity studied, as shown in Table I. $T_c$ decreased and $H_{c2}$ increased as supposed atmospheric exposure increased. Rowell analysis resistivity of the connected portion (which we call adjusted resistivity, $\rho_A$) increased. A less clear trend linking increased atmospheric exposure with a decrease in the calculated current-carrying cross sectional area was also observed. Given that the starting powder was probably exposed to air before it came into our hands and considering that in all cases additional exposure to air during and after milling was rather limited (for example, none of the samples was milled in an unsealed jar) the observed differences are very significant. The remainder of this paper is devoted to explaining the nature of the observed effects.

Key facets of the data that require explanation are as follows: $T_c$ and $H_{c2}$ show a consistent monotonic trend that is consistent with trace alloying exerting intragrain scattering effects. The $T_c$ and $H_{c2}$ trends are generally supported by the measured RRR and normal state resistivity, the latter of which can be controlled by both scattering and by variable sample connectivity. Greater amounts of MgO and $MgB_4$ were found as atmospheric exposure increased.

The x-ray diffraction result indicates larger concentrations of MgO and $MgB_4$ in samples C and D than in samples A and B. This observation is wholly consistent with the milling atmospheres. Exposure to $O_2$ and $H_2O$ appears to have resulted in the reactions

$$2MgB_2 + \tfrac{1}{2}O_2 \rightarrow MgO + MgB_4 \qquad (1)$$

$$2MgB_2 + H_2O \rightarrow MgO + MgB_4 + H_2 \qquad (2)$$

There has been some debate about the line-compound nature of pure $MgB_2$. Although no material is a perfect line compound, the simultaneous increase in $MgB_4$ and MgO content indicate that any formation of Mg vacancies in the $MgB_2$ lattice to compensate for the loss of Mg to MgO must be small. Similarly, these observations imply that oxygen solubility in $MgB_2$ must also be small.

Comparing samples A and D, the ~0.5 K $T_c$ reduction in sample D (the sample with the most exposure to air) can be explained by either a small amount of oxygen doping having a profound effect on $T_c$ or the more likely explanation that the scattering behavior of these samples is changing because carbon present in air as $CO_2$ is being milled into the samples.

Kazakov et al.[25] published a detailed study relating carbon content to $T_c$ in C-alloyed $MgB_2$ single crystals. From their data, one can extract the following relationship for carbon content X < about 0.05:

$$T_c (K) = 39.47 - 82.5 X \qquad (3)$$

Where X is the carbon content expressed by $Mg(B_{1-X}C_X)_2$. Applying eqn. (3) to this sample set and using the $T_c$ values given in Table I (90% normal criterion) we find $T_c(A) - T_c(D) = 0.5$ K. This change in $T_c$ corresponds to $\Delta X \sim 0.006$, or about 0.6% additional substitution of C for B. Applying a modified formalism of the Rowell analysis[15,24,26,27] with $\Delta\rho_{ideal} = 7.3$ $\mu\Omega$-cm we calculated the active area fraction $A_F$, the fraction of the cross section carrying the measurement current, and the adjusted resistivity $\rho_A(40 K)$, the normal state resistivity of this active fraction, finding that $A_F$ ranged between 0.19 and 0.33, which is higher than for the $MgH_2$ - precursor tapes of Matsumoto et al.[28] but less than most of the HIP processed samples in our previous work.[15] As expected, the largest $A_F$ was found for sample A, milled under the cleanest conditions and possessing the least MgO and $MgB_4$, which is consistent with our recent observation that MgO and $MgB_4$ obstruct current flow.[24] Sample A also had the lowest $\rho_A(40 K)$, indicating the smallest amount of electron scattering by defects and dopants. Even though it was the lowest in the set, $\rho_A(40 K) = 2.79$ $\mu\Omega$-cm is quite high compared to pure samples,[19,24] but if we conclude that these samples are C-doped and calculate X from equation (3), then $\rho_A(40 K)$ for these samples agrees rather well with $\rho(40 K)$ measured in fully connected C-doped samples in the literature[25,30] as shown in Fig. 6. The $\rho_A(40 K)$ trend shows an entirely monotonic increase with presumed exposure to air, agreeing well with other indicators of intragranular electron scattering such as $T_c$, $H_{c2}$, and RRR summarized in Table I.

The $A_F$ of Sample B was 12% smaller than sample A, while samples C and D had $A_F$ even lower, linking current obstruction with exposure to air. Curiously, sample D had greater $A_F$ than sample C. We do not have a clear understanding of this facet of the data, but we do stress that both samples C or D have lower $A_F$ than sample A or sample B. Despite recent advances in analysis, an accurate quantitative measure of connectivity is still unavailable.

Further evidence for carbon doping resulting form atmospheric exposure is that for these samples $H_{c2}(T)$ was rather higher than that of the pure filament from Wilke et al. for which $H_{c2}$(24 K, 90% criterion) is only about 6.5 T, but less than the ~ 9 T $H_{c2}$(24 K, 90% criterion) of their X = 0.038 C-doped filament.[30] We found that, as expected, our samples rank in the same order by $dH_{c2}/dT$ as by $\rho_A$(40 K) because $dH_{c2}/dT$ is generally increased by electron scattering. These results are also consistent with our proposal that C is accumulating in the samples as a result of exposure to air.

Low field $J_c(H)$ was found to be similar across our sample set, (samples A, B, C, and D had $J_c$(2 T, 4.2 K) of 212, 244, 220, and 208 kA/cm$^2$ respectively) indicating no strong effect from varied connectivity. More significantly, $J_c(H)$ decreased more slowly with increased magnetic field as the atmospheric exposure increased, with a difference in H*(4.2K) of more than 1 T which means that H*(4.2K) for sample D is about 12% greater than for sample A. This result is consistent with increased carbon doping.

## 5. Conclusions

This experiment studied the effect of variable air exposure during processing on nominally undoped pre-reacted MgB$_2$ subjected to ball milling. Both intergrain and intragrain effects were observed. Small amounts of MgO and MgB$_4$ were observed by x-ray diffraction, indicating reaction with oxygen. The amount of MgO and MgB$_4$ increased with presumed increased exposure to air. The connectivity related parameter A$_F$ was observed to generally decrease as presumed exposure to air increased. All four samples showed greater electron scattering than the purest single crystals and CVD filaments[19,25,30] as indicated by RRR, T$_c$, H$_{c2}$, and $\rho_A$(40 K). Strong circumstantial evidence indicates carbon doping, most notably the similarity between these samples and published data on C-doped CVD filaments.[30] The most likely source of this C is from CO$_2$ in air.

We conclude that $MgB_2$ is highly sensitive to atmospheric $O_2$ and $CO_2$. Reaction with oxygen results in the formation of MgO and $MgB_4$ that could block current flow. Reaction with $CO_2$ dopes $MgB_2$ with carbon, increasing electron scattering and having a strong effect on $T_c$, $H_{c2}$, $H^*$, and resistive properties. We believe these mechanisms are largely responsible for the wide variations in connectivity and differences in $T_c$, resistivity, and $H_{c2}$ reported in the literature for "pure" $MgB_2$.

**Acknowledgements**

This work was supported by the NSF – FRG on $MgB_2$, and by DOE – Understanding and Development of High Field Superconductors for Fusion - DE-FG02-86ER52131. The authors thank their colleagues J. Jiang, W. Starch, and A. Squitieri, in Madison and Tallahassee, and Dr. Yong-Jihn Kim in Mayaguez for discussions.

**Table I** – Processing parameters, designations, and properties.

| Sample | Atmosphere | $T_c$ (SQUID 90%) (K) | $H_{c2}$(24K) (T) | $\rho(40)$ µΩ-cm | $\rho(300)$ µΩ-cm | RRR | $\Delta\rho$ µΩ-cm | $A_F$ | $\rho_A(40)$ µΩ-cm |
|---|---|---|---|---|---|---|---|---|---|
| A | Glove Box | 38.0 | 7.0 | 8.44 | 30.53 | 3.62 | 22 | 0.33 | 2.8 |
| B | Glove Bag | 37.9 | 7.2 | 10.23 | 35.37 | 3.46 | 25 | 0.29 | 3.0 |
| C | Air (sealed) | 37.8 | 7.3 | 17.58 | 56.73 | 3.23 | 39 | 0.19 | 3.3 |
| D | Air (valve) | 37.5 | 7.9 | 14.91 | 41.26 | 2.77 | 26 | 0.28 | 4.1 |

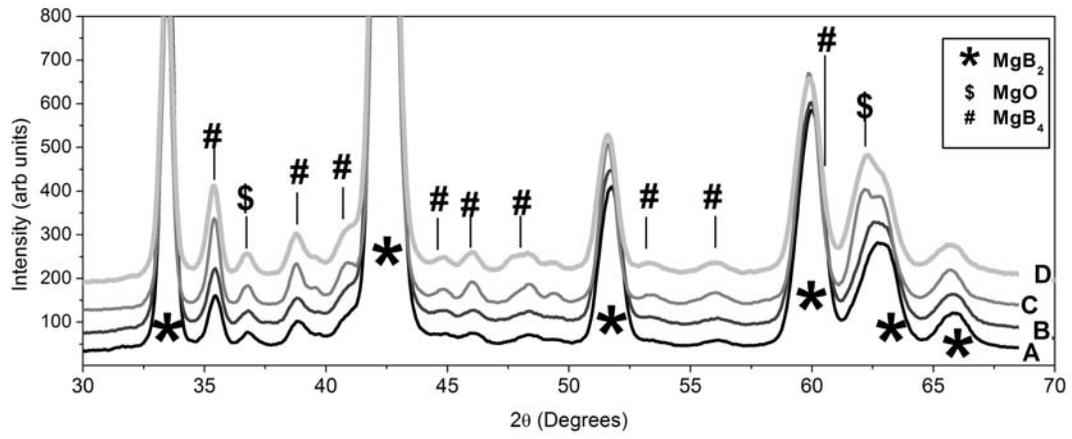

**Figure 1** – X-ray diffraction patterns of heat treated samples. (Cu-K$_\alpha$). Patterns have been offset.

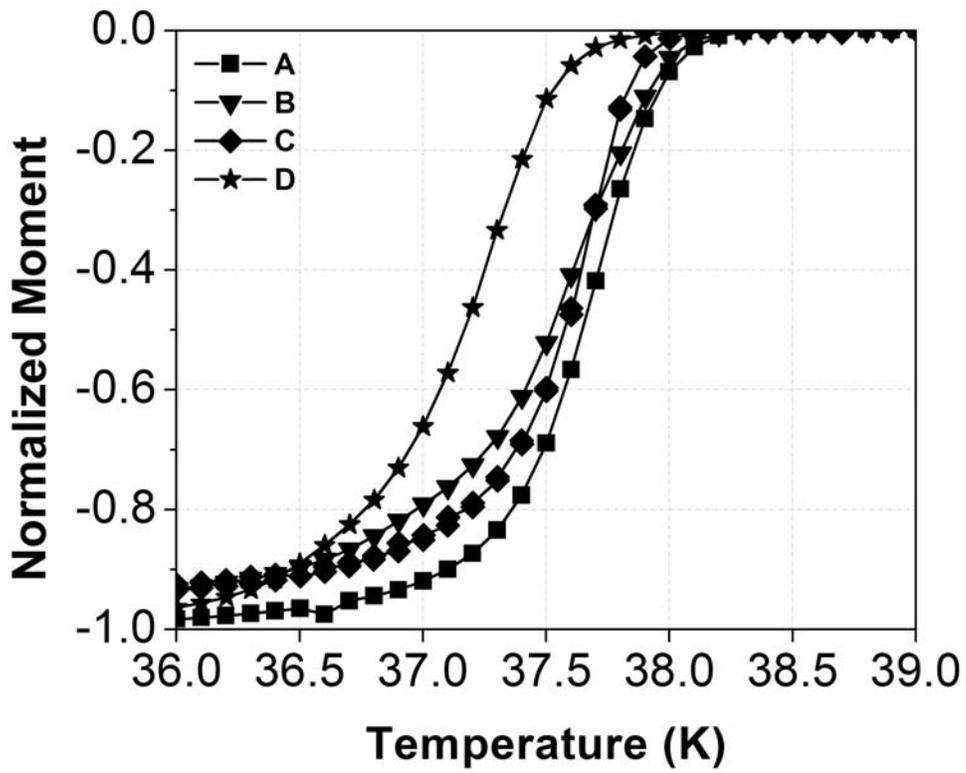

**Figure 2** – Magnetic Tc traces normalized to the magnetic moment at 5 K.

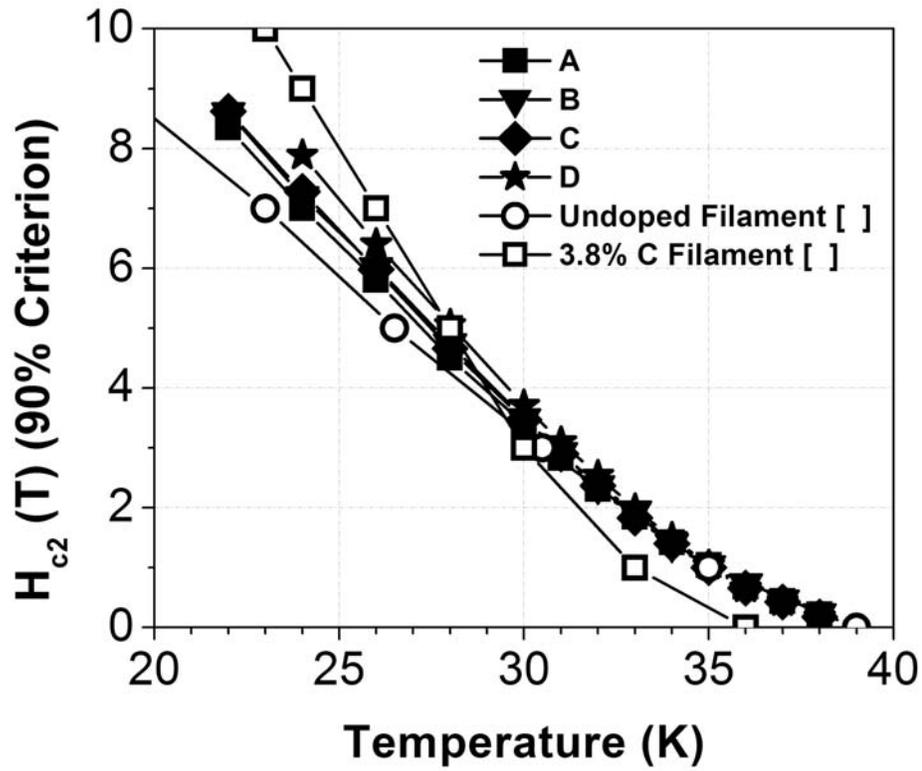

**Figure 3** – $H_{c2}(T)$ using 90% of normal state resistivity criterion. Also included for comparison is data taken from Wilke et al [30] using the same criterion.

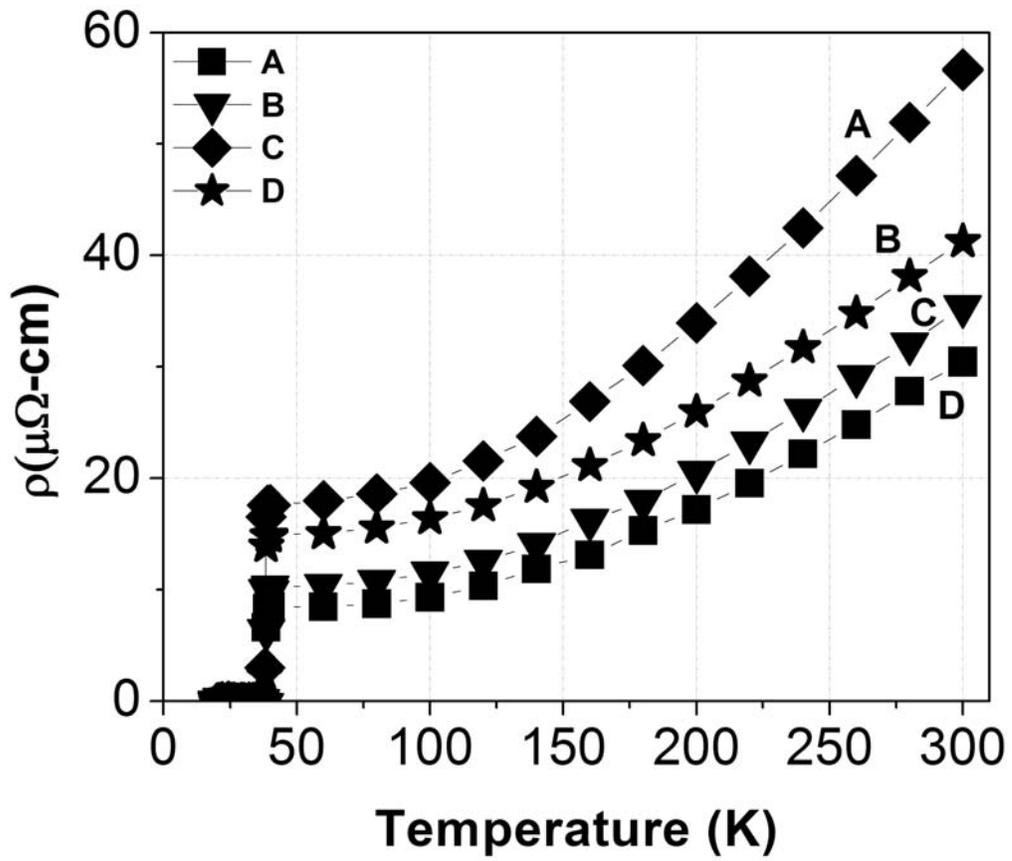

**Figure 4** – $\rho(T)$ from 5 mA, 4-point measurement.

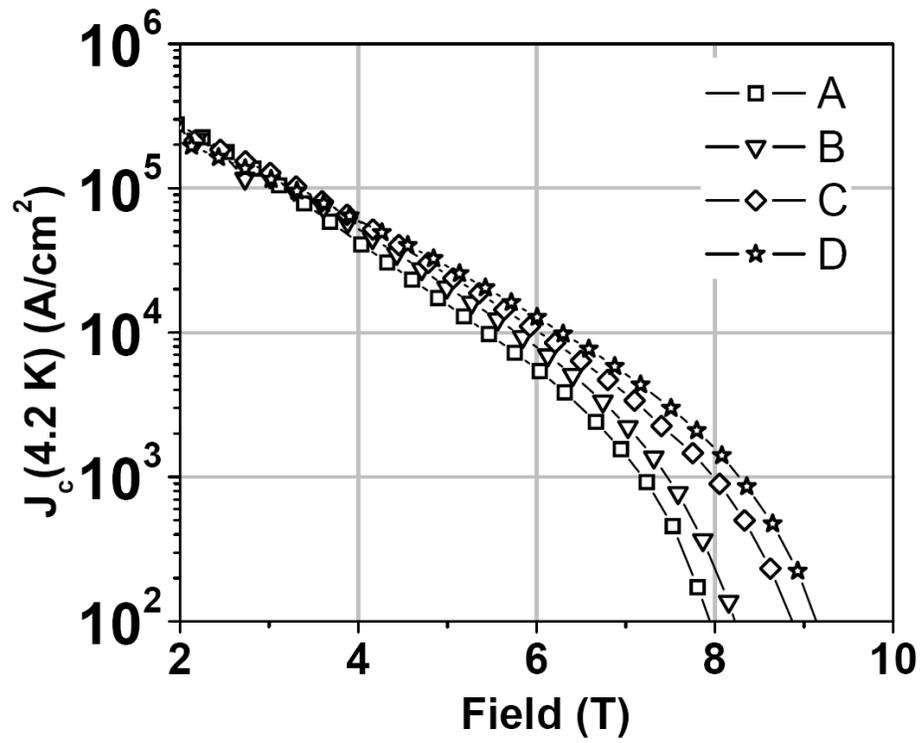

**Figure 5** – $J_c(H)$ at 4.2K determined from MH loops.

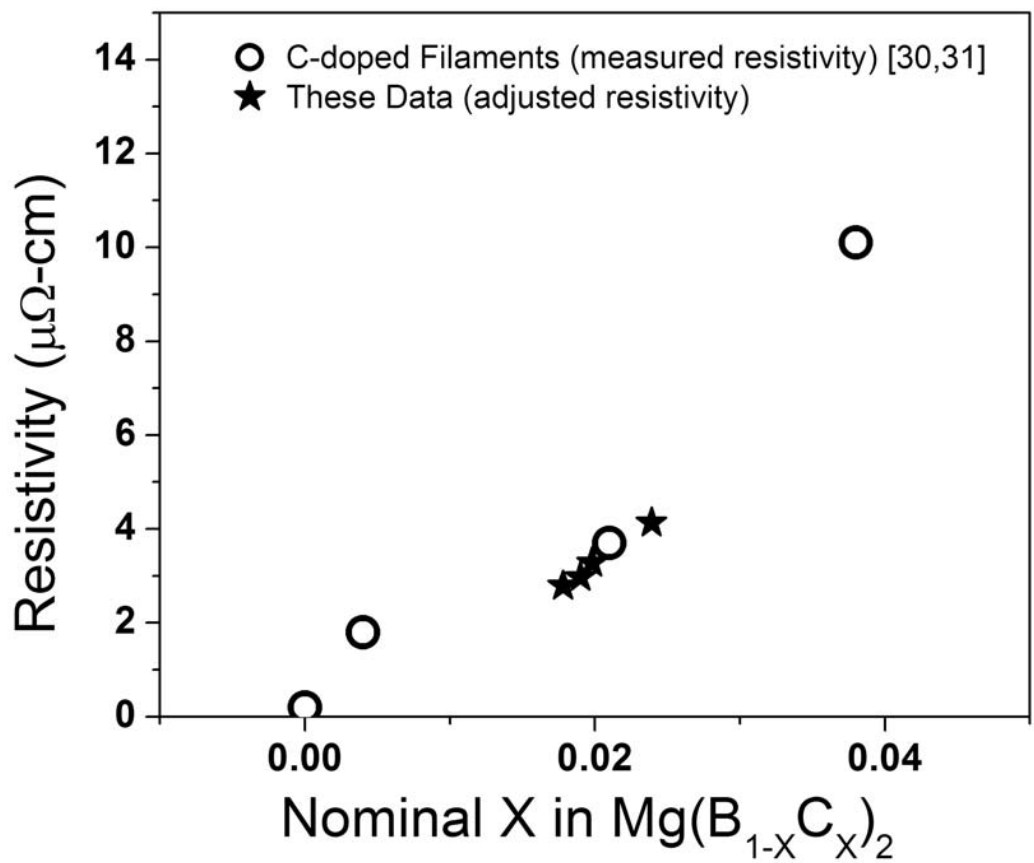

**Figure 6** – Comparison of Rowell adjusted resistivity for these samples to measured resistivity of well-connected C-doped filaments from [30,31]